\input harvmac
\def\lf{16\pi^2}

\def\frak#1#2{{\textstyle{{#1}\over{#2}}}}
\def\frakk#1#2{{{#1}\over{#2}}}
\def\pa{\partial}
\def\semi{;\hfil\break}
\def\ga{\gamma}

\def \th{\theta} 
\def\Xtilde{\tilde X}
\def\Ytilde{\tilde Y}

\def\sic{supersymmetric}
\def\NSVZ{{\rm NSVZ}}
\def\DRED{{\rm DRED}}
\def\DREDp{{\rm DRED}'}

\def\npb{{Nucl.\ Phys.\ }{\bf B}}

\def\prd{{Phys.\ Rev.\ }{\bf D}}

\def\plb{{Phys.\ Lett.\ }{\bf B}}

\def\mtilde{\tilde m}
\def\Ph{\Phi}

\def\th{\theta}
\def\bM{M^*}

\def\bPh{\bar\Ph}
\def\be{\bar\eta}

\def\lf{16\pi^2}
\def\llf{(16\pi^2)^2}
\def\lllf{(16\pi^2)^3}
{\nopagenumbers
\line{\hfil LTH 423}
\line{\hfil hep-ph/9803405}
\vskip .5in    
\centerline{\titlefont The soft scalar 
mass $\beta$-function}
\vskip 1in
\centerline{\bf I.~Jack, D.R.T.~Jones and A.~Pickering}
\medskip
\centerline{\it Dept. of Mathematical Sciences,
University of Liverpool, Liverpool L69 3BX, UK}
\vskip .3in

We present an exact formula for the $\beta$-function for the
soft-breaking scalar 
mass in an $N=1$ supersymmetric gauge theory, in the form
of an operator acting on the  anomalous dimension. In particular we give
the exact form for the correction due to the $\epsilon$-scalar mass, and
show that it has a particularly simple form in the renormalisation
scheme corresponding to the exact NSVZ gauge $\beta$-function.

\Date{ March 1998}}

It has been known for some time that the gauge $\beta$-function
$\beta_g$ for a supersymmetric theory can be expressed (in a
suitable renormalisation scheme) in terms of the anomalous dimension 
matrix $\ga$ of
the chiral superfields, according to the so-called NSVZ
formula\ref\NSVZb{V.~Novikov et al, 
\npb 229 (1983) 381\semi
V.~Novikov et al, \plb166 (1986) 329\semi
M.~Shifman and A.~Vainstein, \npb 277 (1986) 456}. Moreover, for the case of a
softly-broken theory, there has been considerable progress~\ref
\yam{Y.~Yamada, \prd50 (1994) 3537}%
\nref\shif{J.~Hisano 
and M.~Shifman, \prd 56 (1997) 5475}%
\nref\jjg{I.~Jack and
D.R.T.~Jones, \plb 415 (1997) 383}%
\nref\avd{L.V.~Avdeev, D.I.~Kazakov and I.N.~Kondrashuk, \npb510 (1998) 289}%
\nref\jjp{I.~Jack, D.R.T.~Jones and A.~Pickering, hep-ph/9712542}%
--\ref\ahetal{N.~Arkani-Hamed, G.F.~Giudice, M.A.~Luty and 
R.~Rattazzi, hep-ph/9803290}
in writing exact expressions for the $\beta$-functions for the
soft-breaking parameters in terms of $\gamma$.  (The methods of Ref.~\shif,
for incorporating soft-breaking parameters
by replacing couplings by superfields, have 
been extended and used in applications to Beyond the Standard
Model physics, in Refs.~\ahetal\ and \ref\agr{G.F.~Giudice and R.~Rattazzi,
\npb511 (1998) 25}.)
In this paper we discuss the form for the soft-breaking $\phi^*\phi$
mass $\beta$-function; in general this contains a contribution which, at
least when using DRED (\sic\ dimensional regularisation with
minimal subtraction), is related to the
$\epsilon$-scalar mass renormalisation. At present this contribution 
has been calculated explicitly 
only at lowest order in perturbation theory. Here we propose an
exact form for it and present compelling evidence (we believe) in favour of
our result. 

For a $N=1$ supersymmetric gauge theory with superpotential
\eqn\eqf{W(\Phi)={1\over6}Y^{ijk}\Phi_i\Phi_j\Phi_k+
{1\over2}\mu^{ij}\Phi_i\Phi_j,}  
we take the soft breaking Lagrangian $L_{SB}$ as follows:
\eqn\Aba{\eqalign{
L_{SB}(\Ph,W_A)&=-\left\{\int d^2\th\eta\left({1\over6}h^{ijk}\Ph_i\Ph_j\Ph_k
+{1\over2}b^{ij}\Ph_i\Ph_j+{1\over2}MW_A{}^{\alpha}W_{A\alpha}\right)
+{\rm h.c.}\right\}\cr
&\quad -\int d^4\th\be\eta\bPh^j(m^2)^i{}_j(e^{2gV})_i{}^k\Ph_k.\cr}}
Here $\eta=\th^2$ is the spurion external field and $M$ is the gaugino mass.
Use of the spurion
formalism has a long history; in this context see in particular 
Ref.~\ref\yam{Y.~Yamada, \prd50 (1994) 3537}. In \jjg, \avd\
it was shown that $\beta_h$, $\beta_b$ and $\beta_M$ are given by
the following simple expressions:
\eqna\Ai$$\eqalignno{
\beta_h^{ijk}&=\gamma^i{}_lh^{ljk}+\gamma^j{}_lh^{ilk}
+\gamma^k{}_lh^{ijl}-2\gamma_1^i{}_lY^{ljk} 
-2\gamma_1^j{}_lY^{ilk}-2\gamma_1^k{}_lY^{ijl} & \Ai a\cr
\beta_b^{ij}&=\gamma^i{}_lb^{lj}+\gamma^j{}_lb^{il} 
-2\gamma_1^i{}_l\mu^{lj}-2\gamma_1^j{}_l\mu^{il} &\Ai b\cr
\beta_M&={\cal O}\left({\beta_{\alpha}\over \alpha}\right) &\Ai c\cr}$$
where we have written $\alpha=g^2$, 
\eqn\Ajb{
{\cal O}=\left(M\alpha{\pa\over{\pa \alpha}}-h^{lmn}{\pa
\over{\pa Y^{lmn}}}\right),}
and
\eqn\Ajxxx{
(\gamma_1)^i{}_j={\cal O}\gamma^i{}_j.}
The result for $\beta_{m^2}$ is\jjg
\eqn\Ajy{
(\beta_{m^2})^i{}_j=\left[ \Delta + 
\Xtilde(\alpha, Y, Y^*, h, h^*, m, M)\frakk{\pa}{\pa \alpha}\right]
\gamma^i{}_j.}
Here
\eqn\Ajz{
\Delta = 2{\cal O}{\cal O}^* +2M\bM \alpha{\pa
\over{\pa \alpha}} +\Ytilde_{lmn}{\pa\over{\pa Y_{lmn}}}
+\Ytilde^{lmn}{\pa\over{\pa Y^{lmn}}},}
$Y_{lmn} = (Y^{lmn})^*$, and
\eqn\Ajd{
\Ytilde^{ijk}=(m^2)^i{}_lY^{ljk}+(m^2)^j{}_lY^{ilk}+(m^2)^k{}_lY^{ijl}.}
The function $\Xtilde$ introduced in Eq.~\Ajy\ is related to $X$ as defined 
in Ref.~\jjg\ by the equation $\Xtilde = 2gX$. 
The term in $\Xtilde$ does not appear in a naive application of the spurion 
formalism, because (when using DRED) it fails to allow for the fact 
that the $\epsilon$-scalars associated with DRED acquire a mass through
radiative corrections~\ref\jj{I.~Jack and
D.R.T.~Jones, \plb 333 (1994) 372}. Indeed, in DRED, $\beta_{m^2}$ will
actually depend on the $\epsilon$-scalar mass. It is, however, possible to 
define a scheme, $\DREDp$, related to DRED,
such that $\beta_{m^2}$ is independent of the $\epsilon$-scalar mass
\ref\jjmvy{I.~Jack, D.R.T.~Jones,
S.P.~Martin, M.T.~Vaughn and Y.~Yamada, \prd50 (1994) R5481}.
In this scheme $\beta_{m^2}$ is given by Eq.~\Ajy\ with the
leading contribution to $\Xtilde$ given by\jjp 
\eqn\Ajx{\Xtilde = -4S\alpha^2(\lf)^{-1}}
where
\eqn\Awc{
S =  r^{-1}\tr [m^2C(R)] -MM^* C(G).}
(Here $r$ is the number of generators of the gauge group and 
$C(R)$ and $C(G)$ are the quadratic matter and adjoint Casimirs respectively; 
see~\jj\ for more on our notation). In Ref.~\jjp\ we showed
that 
if  there is a RG-invariant 
trajectory $Y=Y(\alpha)$, then  we also have RG-invariant trajectories
given by
\eqna\concc$$\eqalignno{h^{ijk}  &= -2M \alpha\frakk{dY^{ijk}}
{d\alpha} &\concc a\cr
b^{ij} &= -2M\alpha\frakk{d\mu^{ij}}{d\alpha} &\concc b\cr
(m^2)^i{}_j &= 
2\frakk{\alpha^2}{\beta_{\alpha}}MM^*\frakk{d\ga^i{}_j}{d\alpha}, &\concc c\cr
}$$
on which $\Xtilde$ takes the form
\eqn\conca{
\Xtilde = 2\alpha MM^*\left[\frakk{\alpha}{\beta_{\alpha}}
\frakk{d\beta_{\alpha}}{d\alpha}
-2\right].}

Kobayashi et al\ref\kkz{T.~Kobayashi, J.~Kubo and G.~Zoupanos, hep-ph/9802267}
discussed a possible generalisation of our results to a RG-invariant mass sum
rule. They also gave a result for $\Xtilde$ on the RG trajectory in the scheme 
corresponding to the NSVZ results (we shall specify this scheme in more detail 
later). In our conventions, this result
is given by 
\eqn\exX{
\Xtilde^{\NSVZ}=-4{\alpha^2\over{16\pi^2}}
{S\over{\left[1-2\alpha C(G)(16\pi^2)^{-1}\right]}}.}
(In fact, this equation may be obtained by substituting 
into Eq.~\conca\ the NSVZ result for $\beta_{\alpha}$ given by
\eqn\An{
\beta_{\alpha}^{\NSVZ}=2{\alpha^2\over{16\pi^2}}
\left[{Q-2r^{-1}\tr[\gamma C(R)]\over{
1-2\alpha C(G)(\lf)^{-1}}}\right] ,}
where $Q = T(R) - 3C(G)$, and using Eq.~\concc{c}.)
Our principal claim in this paper is that Eq.~\exX\ is
true {\it in general}, and not just on the RG trajectory. 


To substantiate this claim, and to clarify the nature of $\Xtilde$ in 
general, it
is necessary to 
discuss the transformation properties of $\Xtilde$ under a change of
scheme. We consider schemes related by redefinitions 
$\alpha\to \alpha' (\alpha,Y,Y^*)$
and $M\to M'(\alpha, h, M, Y,Y^*)$. It is important not to redefine $Y$ or $h$
if Eqs.~\Ai{}\ are to remain true, and it is easy to convince oneself that 
$m^2$ must not be redefined if Eq.~\Ajy\ is to remain true, which implies
\eqn\delb{\beta'_{m^2}(\alpha',Y,Y^*,h,h^*,m,M')
=\beta_{m^2}(\alpha,Y,Y^*,h,h^*,m,M).}
The transformation properties of $\Xtilde$ then follow from Eq.~\delb.
The first ingredient is the fact that
$\gamma$ and $\gamma_1$ transform in general according to:
\eqna\Ap$$\eqalignno{
\gamma'(\alpha',Y, Y^{*})&=\gamma(\alpha,Y,Y^*),&\Ap a\cr
\gamma_1^{\prime}(\alpha',Y,Y^{*},M',h)
&=\gamma_1(\alpha,Y,Y^*,M,h).&\Ap b\cr}$$
As shown in Ref.~\jjg, Eq.~\Ap{b}\ implies that
\eqn\Ar{
\alpha M
=\alpha^{\prime}M^{\prime}{{\pa \alpha(\alpha',Y,Y^*)}
\over{\pa \alpha'}}
-h^{ijk}{{\pa \alpha(\alpha',Y,Y^*)}\over{\pa Y^{ijk}}}.}
We showed  in Ref.~\jjg\ that Eqs.~\Ai{}\ preserve their form if we make 
scheme redefinitions $\alpha\to \alpha' (\alpha,Y,Y^*)$
and $M\to M'(\alpha, h, M, Y,Y^*)$ as defined above.
It is a straightforward exercise to show 
in a similar fashion that Eq.~\delb\ requires
\eqn\delXa{\eqalign{
\Xtilde= &\Xtilde'{\pa \alpha\over{\pa \alpha'}}+
2M'M^{\prime*}\left[\alpha^{\prime2}{\pa^2\alpha\over{\pa
\alpha^{\prime2}}}
+2\alpha'{\pa\alpha\over{\pa\alpha'}}-2\frakk{\alpha^{\prime2}}{\alpha}
\left(\frakk{\pa\alpha}{\pa\alpha'}\right)^2\right]\cr
&-\left[2M'\alpha'h_{ijk}\left({\pa^2\alpha\over{\pa Y_{ijk}\pa\alpha'}}
-\frakk{2}{\alpha}\frakk{\pa\alpha}{\pa Y_{ijk}}\frakk{\pa\alpha}{\pa\alpha'}
\right)-\Ytilde^{ijk}{\pa\alpha\over{\pa Y^{ijk}}}+\hbox{c.c.}\right] \cr
&+2h^{ijk}h_{lmn}\left[{\pa^2\alpha\over{\pa Y^{ijk}\pa Y_{lmn}}}
-\frakk{2}{\alpha}\frakk{\pa\alpha}{\pa Y^{ijk}}
\frakk{\pa\alpha}{\pa Y_{lmn}}\right]
,\cr}}
which may be rewritten more succinctly as
\eqn\delX{
\Xtilde=\Xtilde'{\pa \alpha\over{\pa \alpha'}}+(\Delta'-\Delta)\alpha,}
where $\Delta$ is given in Eq.~\Ajz.
Reassuringly, one can then check that on an RG-invariant
trajectory, Eq.~\conca\ is true 
in any scheme related by $\alpha\to \alpha' (\alpha,Y,Y^*)$
and $M\to M'(\alpha, h, M, Y,Y^*)$ (as in Eq.~\Ar). Indeed, 
on an RG-invariant trajectory, we
find 
\eqn\tnfn{\eqalign{
\beta'_{\alpha'}&= {d\alpha'\over{d\alpha}}\beta_{\alpha}\cr
\alpha'M'&=\alpha M{d\alpha'\over{d\alpha}}\cr}}
(using Eqs.~\concc{a}\ and \Ar).  
Then if $\Xtilde$ is given in the unprimed scheme by Eq.~\conca,
it follows from Eq.~\delXa\ 
that $\Xtilde'$ satisfies the same equation in the primed scheme. 

The $\DREDp$ scheme is well-defined; but what we mean by the NSVZ
scheme  requires further explanation. We have shown\ref\jjn{I.~Jack,
D.R.T.~Jones and  C.G.~North, \npb486 (1997) 479} how to construct
perturbatively a redefinition  $\alpha^{\DRED}\to\alpha^{\NSVZ}$ which
takes us from DRED to a scheme in which $\beta_{\alpha}$  takes the NSVZ
form of Eq.~\An.
We now define the NSVZ scheme as the result of making this redefinition of 
$\alpha$, accompanied by the corresponding redefinition of $M$ as given in 
Eq.~\Ar. We believe that this scheme corresponds to that used by Hisano and 
Shifman\shif\ to derive their exact results for the soft-breaking couplings. 
This is because their scheme certainly corresponds to $\beta_{\alpha}$ of the 
form Eq.~\An, and moreover their result (consisting of an RG-invariant 
combination of $M$, $b$ 
and $\alpha$) could be shown to lead to our exact formula for $\beta_M$ in 
Eq.~\Ai{c}. The scheme used by Hisano and Shifman is defined by changing from
the holomorphic normalisation for the gauge coupling in the Wilsonian action
to the canonical normalisation. In this context, it is of interest to consider
the reverse transition, from the NSVZ scheme to the holomorphic scheme. 
(For a discussion of the relationship between the holomorphic and NSVZ 
couplings see 
Ref.~\ref\ahhm{N.~Arkani-Hamed and H.~Murayama, hep-th/9707133}.)
Consider a theory with only a single Yukawa coupling $y$ and with  
both $C(R)$ and $\ga$ 
proportional to the unit matrix. For an example of such a theory, 
consider $SU(N)$ with three adjoint 
chiral superfields and the superpotential 
\eqn\nfourd{W= yd^{abc} \phi_1^a\phi_2^b\phi_3^c.} 
Let us define the transformation 
\eqn\hol{
{1\over{\alpha^{{\rm H}}}}={1\over{\alpha^{\NSVZ}}}+\frakk{2}{\lf}
C(G)\ln\alpha^{{\NSVZ}}
-{2\over3}{1\over{\lf}}T(R)\ln(yy^*),}
with an associated transformation of $M$ dictated by Eq.~\Ar. 
In the ``holomorphic'' scheme corresponding to $\alpha^{{\rm H}}$,
$\beta_{\alpha^{{\rm H}}}$ is one-loop exact, i.e. $\lf\beta_{\alpha^{{\rm H}}}
=2Q(\alpha^{{\rm H}})^2$. 
(In our example Eq.~\nfourd, we have, in fact,  $Q=0$ and so in this case 
$\beta_{\alpha^{{\rm H}}}$ vanishes to all orders). 
Remarkably, we find using Eq.~\delXa\ that $\Xtilde^{{\rm H}}=0$,
assuming that $\Xtilde^{\NSVZ}$ is given by Eq.~\exX. This means that in the
holomorphic scheme, all contributions to the soft $\beta$-functions 
beyond one loop can be traced to $\gamma$. 
The fact that the form of $\Xtilde^{\NSVZ}$ 
leads to the vanishing of $\Xtilde^{{\rm H}}$ is
intriguing. In a more general case with 
$C(R), \ga$ not proportional to the unit matrix it is straightforward
to write down a generalisation of Eq.~\hol\ such that $\beta_{\alpha^H}$ 
is one-loop
exact; we have simply to replace $\frak{2}{3}T(R)\ln(yy^*)$ by 
$4r^{-1}\tr [ZC(R)]$ where $\mu\frakk{dZ}{d\mu} = \ga$. 
However, we do not then find that $\Xtilde^{{\rm H}}$ is zero, 
except of course on an RG-invariant trajectory, Eq.~\conca.

Our proposed exact result for $\Xtilde$ is in the NSVZ scheme; however, 
the existence of $\Xtilde$ was first identified in the $\DREDp$ scheme,
and ascribed to the $\epsilon$-scalar mass. In fact, we shall also
derive an exact formula for $\Xtilde$ in $\DREDp$, related to the
$\beta$-function  $\beta_{\mtilde^2}$ for the $\epsilon$-scalar mass
$\mtilde^2$. We shall then show that our proposed result for $\Xtilde$ in 
the NSVZ scheme is related to our result for $\Xtilde$ in $\DREDp$ by 
Eq.~\delXa, up to the limits of our perturbative calculations.

We now give our result for $\Xtilde$ in $\DREDp$.
Suppose that $\beta_{\mtilde^2}$ is given by
\eqn\betatm{
\beta_{\mtilde^2}=N_1+N_2\mtilde^2,}
where $N_1(\alpha,Y,Y^*,h,h^*,m,M)$ 
does not depend on $\mtilde$, and may be written
\eqn\betatma{
N_1=\sum_{L=1}^{\infty}N_1^{(L)},}
where $N_1^{(L)}$ is the $L$-loop contribution to $N_1$. Our claim now is
that
\eqn\Xdred{
\Xtilde^{\DREDp} =-\sum_{L=1}^{\infty}{\alpha\over L}N_1^{(L)}.}
The factor of $1/L$ arises here because $\Xtilde$ is related to the 
{\it renormalisation constant\/} that determines $\beta_{\mtilde^2}$, which 
differs at $L$ loops by a factor of $L$ from $\beta_{\mtilde^2}$ itself.  
The proof of Eq.~\Xdred\ is as follows:
$\DRED$ and $\DREDp$ are related by a redefinition of $m^2$ given by
\eqn\delm{
(m^2)^{\DREDp}=(m^2)^{\DRED}+\delta m^2,}
where\ahetal
\eqn\delma{
\delta m^2=
\mtilde^2\sum_{L=1}^{\infty}{1\over{L}}\alpha{\pa\over{\pa\alpha}}\gamma^{(L)}.}
$\delta m^2$ was computed at lowest order in Refs.~\jj, \ref\mv{S.P.~Martin
and M.T.~Vaughn, \prd50 (1994) 2282}, and \jjmvy, and 
corresponds\jj\ref\jjr{I.~Jack, D.R.T.~Jones and K.L. Roberts,
Z.~Phys. {\bf C}63 (1994) 151}\ 
to finite contributions to $(m^2)^B$ which contain 
a simple pole in $\epsilon$ multiplied by a factor of $\epsilon$ deriving 
from a loop of $\epsilon$-scalars, and containing an insertion of the 
$\epsilon$-scalar mass. The rules originally derived by Yamada\yam\ from the
spurion formalism lead to Eq.~\Ajy\ without the term in $\Xtilde$. From 
a diagrammatic point of view, these rules precisely omit (compared to DRED)
diagrams
with an $\epsilon$-scalar mass counterterm replacing $\mtilde^2$ 
in a diagram contributing to $\delta m^2$. In particular, the missing simple 
pole terms (which account for the difference between 
$\beta_{m^2}^{\DRED}$ and $\Delta\ga$, where $\Delta$ is defined in Eq.~\Ajz) 
correspond to replacing 
$\mtilde^2$ in $\delta m^2$ by the simple pole in $(\mtilde^2)^B$. So we find 
\eqn\delmb{
\beta_{m^2}^{\DRED(L)}=\Delta\ga^{(L)}-\sum_{r=1}^{L-1}L{N_1^{(L-r)}
\over{L-r}}
{1\over{r}}\alpha{\pa\over{\pa\alpha}}\gamma^{(r)}+{\rm O}(\mtilde^2).}
The factor of $L$ is required to convert the simple pole residue 
to the $\beta$-function. Moreover, we also have
\eqn\delmc{
\beta_{m^2}^{\DREDp}(m^2+\delta m^2)=\beta_{m^2}^{\DRED}(m^2)+
\mu{\pa\over{\pa \mu}}\delta m^2,}
which implies 
\eqn\delme{
\beta_{m^2}^{\DREDp(L)}=\beta_{m^2}^{\DRED(L)}+\sum_{r=1}^{L-1}N_1^{(L-r)}
{1\over{r}}\alpha{\pa\over{\pa\alpha}}\gamma^{(r)}+{\rm O}(\mtilde^2).}
Combining Eqs.~\delmb\ and \delmc, we have
\eqn\delmd{
\beta_{m^2}^{\DREDp(L)}=\Delta\ga^{(L)}-\sum_{r=1}^{L-1}{N_1^{(L-r)}
\over{L-r}}\alpha{\pa\over{\pa\alpha}}\gamma^{(r)},}
which, with Eq.~\Ajy, implies Eq.~\Xdred. (In Eq.~\delmd\ we have assumed 
the cancellation of the terms in $\mtilde^2$ in $\beta_{m^2}^{\DREDp}$; this 
was shown up to two loops in Ref.~\jjmvy, and has recently been proved to all
orders in Ref.~\ahetal.)

We can corroborate this claim for $\Xtilde^{\DRED}$ by 
explicit perturbative calculations.
The $\epsilon$-scalar mass $\beta$-function was calculated in Ref.~\jj\ up to
two loops; from which result it is easy to obtain $N_1^{(1)}$  and 
$N_1^{(2)}$. From Eq.~\Xdred\ we then obtain
\eqn\Xpert{\eqalign{
\lf \Xtilde^{(1)}&=-4\alpha^2S,\cr
\llf \Xtilde^{\DREDp(2)}&=r^{-1}\alpha^2\tr [ W C(R)]
-8\alpha^3C(G)S-4\alpha^3C(G)QMM^*,\cr}}
where
\eqn\Wdef{\eqalign{
W^j{}_i&={1\over2}Y_{ipq}Y^{pqn}(m^2)^j{}_n+{1\over2}Y^{jpq}Y_{pqn}(m^2)^n{}_i
+2Y_{ipq}Y^{jpr}(m^2)^q{}_r\cr &\quad
+h_{ipq}h^{jpq}-8\alpha MM^*C(R)^j{}_i.\cr}}
(We note that the expression for $\beta_{\mtilde^2}^{(2)}$ 
in Ref.~\jj\ contained a factor of $2$ 
misprint in the  coefficient of $\alpha^3C(G)S$ which we have corrected
here.)  Clearly $\Xtilde^{(1)}$ agrees with Eq.~\Ajx, 
and indeed with the lowest order contribution to
$\Xtilde^{\NSVZ}$ in Eq.~\exX. From Eq.~\exX, we also have 
\eqn\Xtwo{
\llf \Xtilde^{\NSVZ(2)}= -8\alpha^3C(G)S.}
To compare $\Xtilde^{\DREDp(2)}$ and $ \Xtilde^{\NSVZ(2)}$, we need to know the
coupling constant  redefinition linking NSVZ and DRED.  Writing
\eqn\redefa{
\alpha^{\DRED}= \alpha^{\NSVZ}+\sum_{L=1}^{\infty}
\delta^{(L)}(\alpha^{\NSVZ},Y,Y^*),}
where $\delta^{(L)}(\alpha^{\NSVZ},Y,Y^*)$ is the $L$-loop redefinition, 
it was shown in Ref.~\jjn\ that 
$\delta^{(1)}=0$ and 
\eqn\tlfbb{
\llf\delta^{(2)}=\alpha^2
\left[r^{-1}\tr\left\{PC(R)\right]-\alpha QC(G)\right\},}
where
\eqn\defs{
P^i{}_j=\frak{1}{2}Y^{ikl}Y_{jkl}-2\alpha C(R)^i{}_j. }
It is easy to verify using Eq.~\delXa\ that $\Xtilde^{\NSVZ(2)}$ as given in
Eq.~\Xtwo\ transforms into $ \Xtilde^{\DREDp(2)}$ as given in Eq.~\Xpert\ when
$\alpha$ transforms according to Eq.~\tlfbb. We have also performed a
partial three-loop calculation of $\beta_{\mtilde^2}$ in order to provide an
even more stringent check. We have computed the contributions with a
maximal number of Yukawa couplings, of the form $\alpha Y^2Y^{*2}m^2$.
We performed the  calculation in components, using the $\epsilon$-scalar
lagrangian which may be  found in Ref.~\ref\cjv{D.M.~Capper, D.R.T.~Jones
and P.~van Nieuwenhuizen, \npb167 (1980) 479}. The computation presents no
special difficulties, since the  momentum integrals may be reduced to
the basic set given in Ref.~\jjn, and we  suppress the details. We find
a prediction for the three-loop contribution to $\Xtilde^{\DREDp}$ 
which we can write in the form
\eqn\Xthreea{\eqalign{
\lllf\Xtilde^{\DREDp(3)}=&\alpha^2\left(\Ytilde^{ijk}{\pa\over{\pa Y^{ijk}}}
+\Ytilde_{ijk}{\pa\over{\pa
Y_{ijk}}}\right)\Bigl({1\over{12r}}\tr(Y^2Y^2C(R))\cr
&-{2\over{3r}}Y^{ijk}Y_{ilm}(Y^2)^l{}_jC(R)^m{}_k\Bigr)+\ldots.\cr}}
Since we have from Eq.~\exX\ that $\lllf\Xtilde^{\NSVZ(3)}
= -16\alpha^4[C(G)]^2S$, 
all the terms given explicitly in Eq.~\Xthreea\ must be
generated by the transformation from NSVZ to $\DREDp$. 
It was shown in Ref.~\jjn\ that
\eqn\redefb{
\delta^{(3)}=\rho\Delta_1-{4\over3}\Delta_2+{1\over3}\Delta_3}
where 
\eqna\fourb$$\eqalignno{ \lllf\Delta_1&=
\alpha^3C(G)\left[r^{-1}\tr[PC(R)] - \alpha QC(G)\right]&\fourb a\cr
\lllf\Delta_2&= r^{-1}\tr\left[\alpha^2S_4C(R) -2 \alpha^4QC(R)^2 + 2
\alpha^3PC(R)^2\right]  &\fourb b\cr \lllf\Delta_3&=
\alpha^2r^{-1}\tr[P^2C(R)] - \alpha^4 Q^2C(G). &\fourb c\cr}$$ 
Here $S_4 ^i{}_j = Y^{imn}P^p{}_m Y_{jpn}$. 
(The
coefficient $\rho$ has not yet been calculated explicitly\foot{In 
Ref.~\ref\jjs{I.~Jack, D.R.T.~Jones and M.A.~Samuel, \plb 407 (1997) 143.} 
a method based on Pad\'e approximants 
was used to suggest that $\rho\approx 4.8$}, 
but it is immaterial
for our present purposes.) It is straightforward to show that this
transformation generates the terms in Eq.~\Xthreea\ according to
Eq.~\delXa. This completes our claim that $\Xtilde$ is given 
by Eq.~\exX\ in NSVZ and by Eq.~\Xdred\ in $\DREDp$. 

The authors of 
Ref.~\ahetal\ present an elegant prescription for accommodating $\DREDp$ within
the formalism of Ref.~\shif. 
This leads to a method for extracting $\beta_{m^2}$ which should 
lead to results comparable to ours, though they do not give an explicit 
all-orders formula.

Given our result for $\Xtilde$, we now have a complete set of exact 
$\beta$-function results in the NSVZ scheme for an arbitrary 
gauge theory. All the $\beta$-functions are given in terms of the 
anomalous dimension matrix $\ga$ of the chiral superfields. Since 
$\ga^{\NSVZ}$ has been given through three loops in Ref.~\jjn\ there 
is no obstacle to, for example,  performing the usual MSSM running 
analysis through three loops. Results valid in $\DREDp$ 
could then be obtained 
at $M_Z$ (say) by using the redefinition of $\alpha$ as given in 
Eqs.~\redefa, \tlfbb\ and \fourb{}\ and 
the redefinition of $M$  given in Eq.~\Ar.

\bigskip\centerline{{\bf Acknowledgements}}\nobreak
 
AP was supported by a PPARC Research Grant.
 
\listrefs

\bye